\documentclass[twocolumn,prl,aps,amsmath,amssymb,floatfix,showpacs]{revtex4}
\usepackage{graphicx}
\usepackage{bm}

\begin{document}
\title{Superconductivity in C$_6$Ca explained. }
\author{Matteo Calandra}
\author{Francesco Mauri}
\affiliation{Institut de Min\'eralogie et de Physique des Milieux condens\'es, 
case 115, 4 place Jussieu, 75252, Paris cedex 05, France}
\date{\today}

\begin{abstract}
Using density functional theory we demonstrate that
superconductivity in C$_6$Ca is due to a phonon-mediated mechanism with
electron-phonon coupling $\lambda=0.83$ and phonon-frequency logarithmic-average 
$\langle \omega \rangle=24.7$ meV. 
The calculated isotope exponents are
$\alpha({\rm Ca})=0.24$ and $\alpha({\rm C})=0.26$.
Superconductivity is mostly due C 
vibrations perpendicular and Ca vibrations parallel to the graphite layers.
Since the electron-phonon couplings of these modes are activated by the presence of an 
intercalant Fermi surface, the occurrence of superconductivity in 
graphite intercalated compounds requires a non complete ionization of the 
intercalant.

\end{abstract}
%%      71.15.Mb Density functional theory, local density approximation, 
%%               gradient and other corrections
%%      71.20.Tx Fullerenes and related materials; intercalation compounds
%%               Superconductivity:
%%      74.25.Jb Electronic structure
%%      74.25.Kc Phonons
%%      74.70.-b Superconducting materials  (for cuprates see 74.72.-h)
%%      74.70.Ad Metals; alloys and binary compounds (including A15, MgB2, etc.)
\maketitle

Graphite intercalated compounds (GICs) were first synthesized in 1861 \cite{Schaffautl}
but only from the 30s a systematic study of these systems began. Nowadays a large number of
reagents can be intercalated in graphite ($\gg 100$)\cite{DresselhausRev}. 
Intercalation allows to change continuously the properties 
of the pristine graphite system, as it is the case for the electrical conductivity. 
The low conductivity of graphite can be enhanced to obtain even larger
conductivities than Copper \cite{Foley}. Moreover at low temperatures, 
intercalation can stabilize a superconducting state\cite{DresselhausRev}.
The discovery of superconductivity in other intercalated structures like
MgB$_2$\cite{Nagamatsu} and in other forms of doped Carbon (diamond)
\cite{Ekimov} has renewed interest in the field. 

The first discovered GIC superconductors  were alkali-intercalated
compounds\cite{Hannay} (C$_8$A with A= K, Rb, Cs with T$_c <$ 1 K).  
Synthesis under pressure has been used to obtain metastable GICs 
with larger concentration of alkali
metals (C$_6$K, C$_3$K, C$_4$Na, C$_2$Na) where the highest T$_c$
corresponds to the largest metal concentration, T$_c$(C$_2$Na)=5 K \cite{Belash}.
Intercalation involving several stages have also been
shown to be superconducting\cite{Alexander,Outti} (the highest 
T$_c$ = 2.7 K in this class belongs to KTl$_{1.5}$C$_{4}$).
Intercalation with rare-earths has been tried, 
C$_6$Eu, C$_6$Cm and C$_6$Tm are not superconductors,
while  recently it has been shown that C$_6$Yb has a T$_c$ = 6.5 K \cite{Weller}. 
Most surprising superconductivity on a non-bulk sample of C$_6$Ca was also 
discovered\cite{Weller}. The report
was confirmed by measurements on bulk C$_6$Ca poly-crystals\cite{Genevieve} and a 
$T_c=11.5$ K was clearly identified. At the moment C$_6$Yb and C$_6$Ca are the GICs with 
the highest T$_c$. It is worthwhile to remember that elemental Yb and Ca 
are not superconductors.

Many open questions remain concerning the origin of superconductivity in 
GICs. 
(i) All the aforementioned intercalants act as donors respect to graphite but 
there is no clear trend between the number of carriers transferred
to the Graphene layers and T$_c$\cite{DresselhausRev}. 
What determines T$_c$?
(ii) Is superconductivity due to the electron-phonon interaction \cite{Mazin} or
to electron correlation \cite{Csanyi}? 
(iii) In the case of a phonon mediated pairing which are the relevant phonon
modes \cite{Mazin}? 
(iv) How does the presence of electronic donor states (or interlayer states) 
affect superconductivity \cite{DresselhausRev,Csanyi,Mazin}?

\begin{figure*}[t]
\rotatebox{90}{\includegraphics[height=5.17cm]{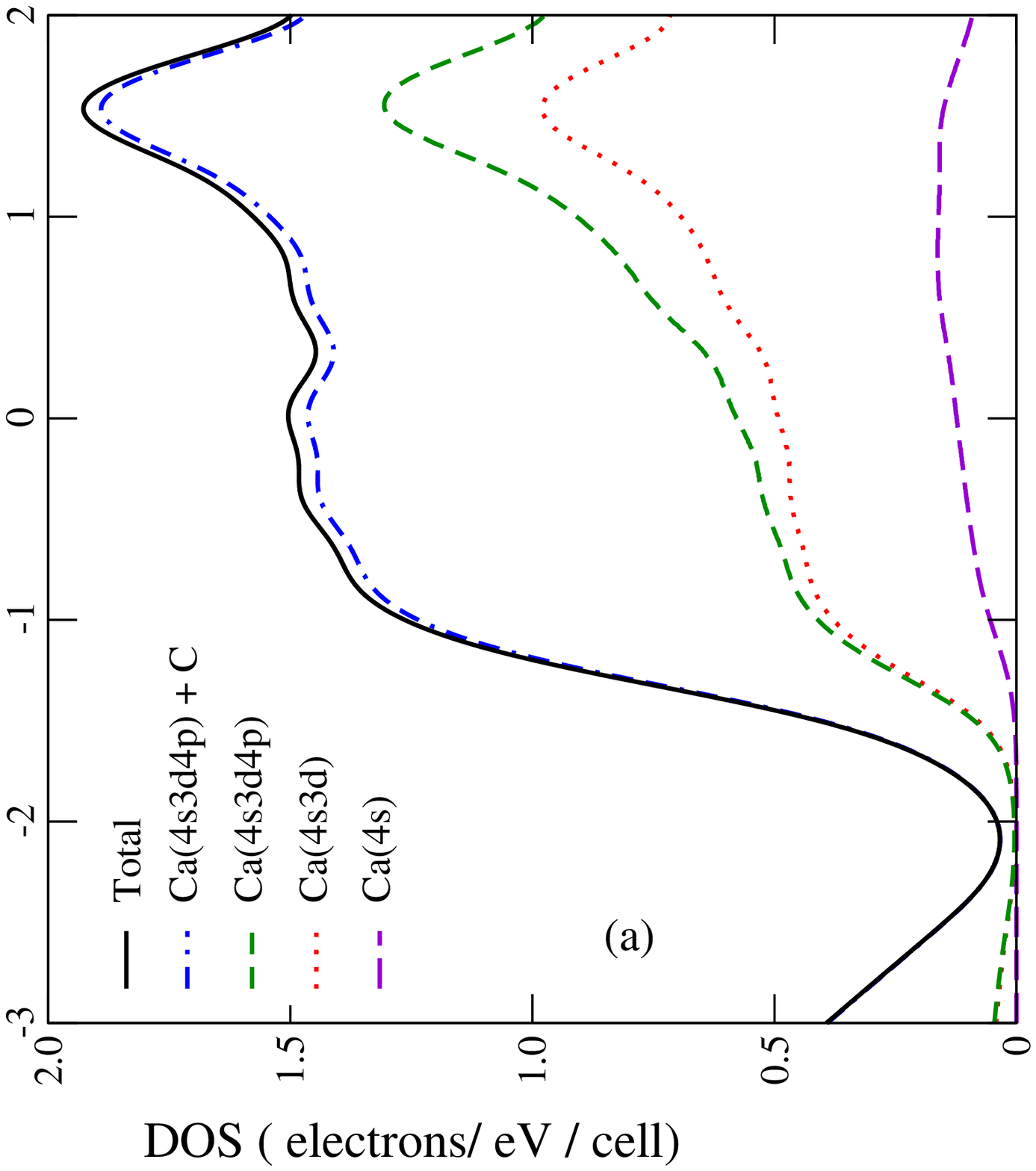}}%
\includegraphics[height=5.5cm]{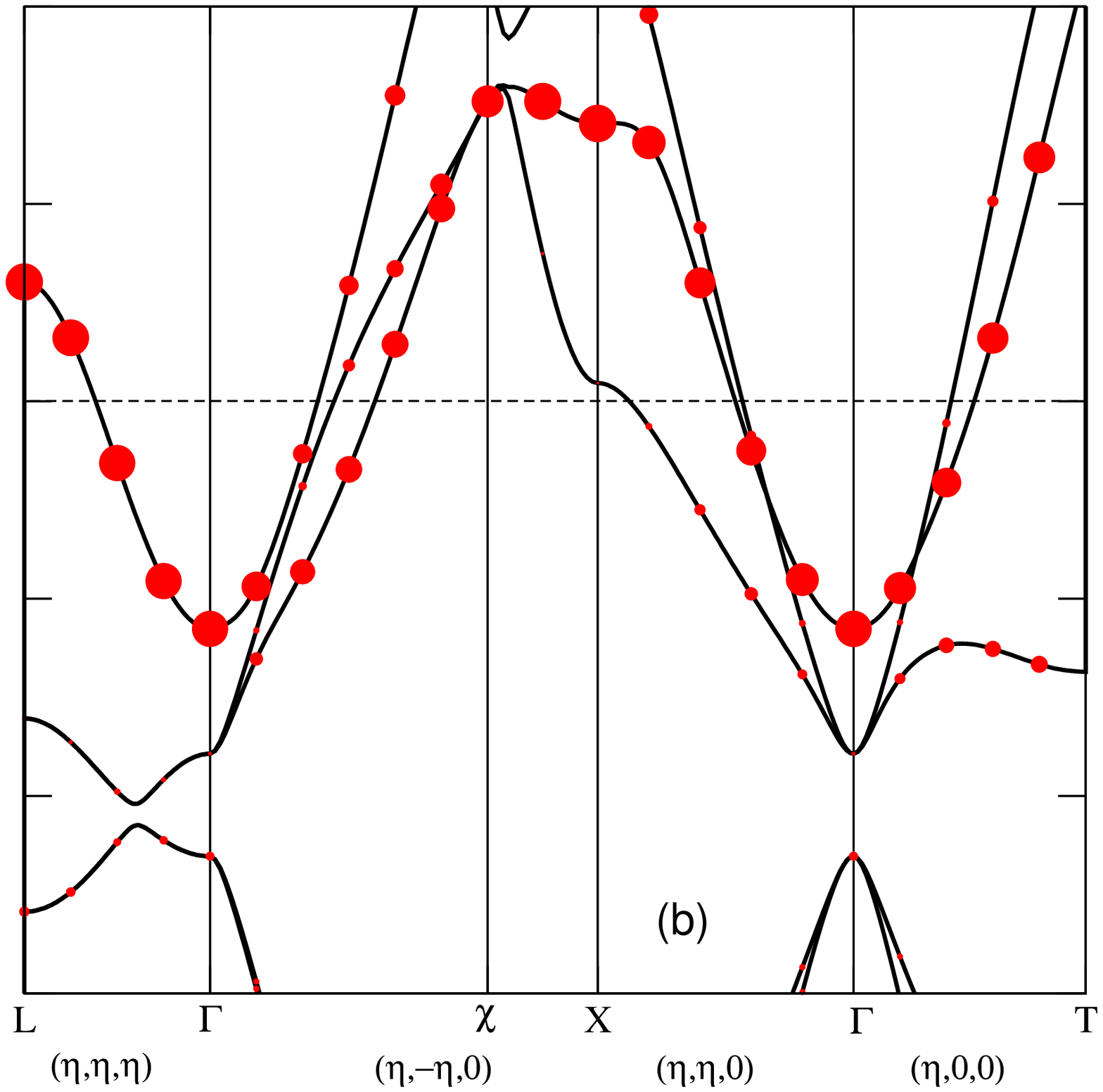}%
\includegraphics[height=5.5cm]{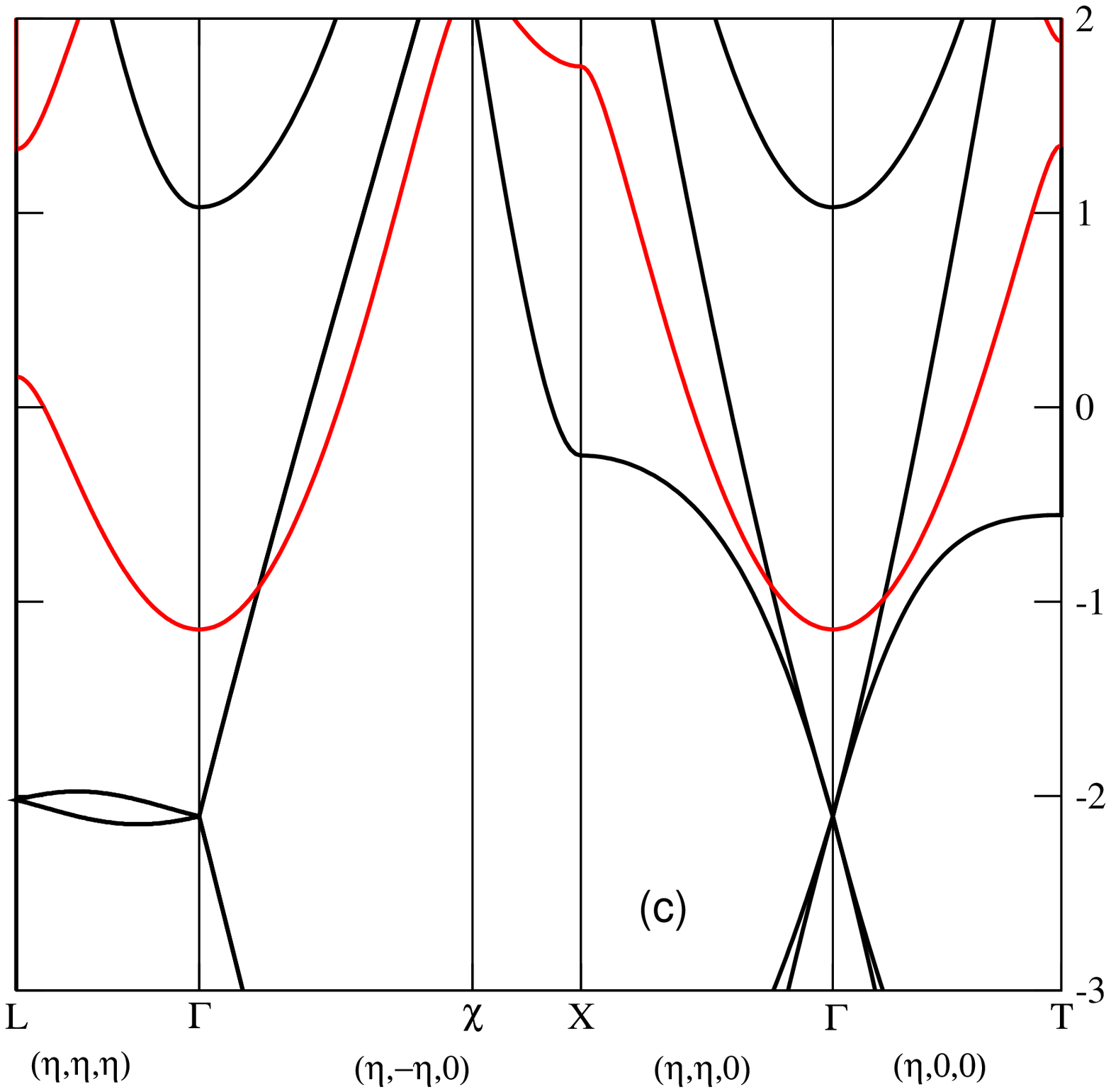}%
\caption{(Color online) (a) Total DOS and DOS projected on selected atomic 
wavefunctions in C$_6$Ca. (b) Band structure of C$_{6}$Ca. The size of the dots 
represents the percentage of Ca component. As a reference
the dot at $\approx 0.6$ eV  at the L-point represents 0.95 Ca component. (c)
Band structure of $^*$Ca (Red) and C$_6^*$ (black). The bands have been shifted
to compare with the C$_6$Ca band structure. 
The directions are given in terms of the rombohedral reciprocal lattice vectors.
$\chi$ is the interception between the $(\eta,-\eta,0)$ line with the border 
of the first Brillouin zone ($\eta=0.3581$). The points L, $\Gamma$, X, T have $\eta=0.5$.}
\label{fig:bands}
\end{figure*}
Two different theoretical explanations has been proposed for superconductivity 
in C$_6$Ca. In \cite{Csanyi} it was noted that in most superconducting GICs an interlayer
state is present at E$_f$ and a non-conventional excitonic 
pairing mechanism\cite{Allender} has been proposed. On the contrary Mazin \cite{Mazin} 
suggested  an ordinary electron-phonon pairing mechanism involving mainly
the Ca modes with a 0.4 isotope exponent for Ca and 0.1 or less for C. 
However this conclusion is not based on calculations of
the phonon dispersion and of the electron-phonon coupling in C$_6$Ca.
Unfortunately isotope measurements supporting or
discarding these two thesis are not yet available.

In this work we identify unambiguously the mechanism responsible for superconductivity 
in C$_6$Ca. Moreover we calculate the phonon dispersion
and the electron-phonon coupling. We predict the values
of the isotope effect exponent $\alpha$ for both species.

We first show that the doping of a graphene layer and an electron-phonon 
mechanism cannot explain the observed T$_c$ in superconducting GICs.
We assume that doping acts as a rigid shift of the graphene Fermi level.
Since the Fermi surface is composed by $\pi$ electrons, which are antisymmetric respect
to the graphene layer, the  out-of-plane phonons do not contribute to the 
electron-phonon coupling $\lambda$. 
At weak doping $\lambda$ due to in-plane phonons can be computed 
using the results of ref. \cite{Piscanec} . 
The band dispersion can be
linearized close to the K point of the hexagonal structure,
and the 
density of state 
per two-atom graphene unit-cell is $N(0)=\beta^{-1}\sqrt{8\pi\sqrt{3}}\sqrt{\Delta}$
 with $\beta=14.1$ eV and $\Delta$ is the number of electron donated per unit cell (doping).
Only the E$_{2g}$ modes near $\Gamma$ and 
the A$^{\prime}_{1}$ mode near K contribute:
\begin{equation}\label{eq:model}
\lambda=N(0)\left[
\frac{2\langle g^{2}_{\bf \Gamma}\rangle_{F}}{\hbar \omega_{\bf \Gamma}}+
\frac{1}{4}\frac{2\langle g^{2}_{\bf K}\rangle_{F}}{\hbar \omega_{\bf K}}\right]=0.34\sqrt{\Delta}
\end{equation}
where the notation is that of ref. \cite{Piscanec}. Using this equation and typical values of $\Delta$ 
\cite{Pietronero}
the predicted T$_c$ are order of magnitudes 
smaller than those observed. As a consequence superconductivity in C$_6$Ca and in 
GICs cannot be simply interpreted as doping of a graphene layer, 
but it is necessary to consider the GIC's full structure.

The atomic structure\cite{Genevieve} of CaC$_{6}$ involves a stacked arrangement 
of graphene sheets (stacking AAA) with Ca atoms occupying interlayer sites above 
the centers of the hexagons (stacking $\alpha\beta\gamma$). 
The crystallographic structure is R\={3}m \cite{Genevieve} where 
the Ca atoms occupy the 1a Wyckoff position (0,0,0) and the C atoms 
the 6g positions (x,-x,1/2)
with x$=1/6$. 
The rombohedral elementary 
unit cell has 7 atoms,  lattice parameter  5.17 ${\rm \AA}$ and 
rombohedral angle $49.55^o$. 
The lattice formed by Ca atoms in C$_6$Ca can be seen as a deformation of that 
of bulk Ca metal. Indeed the fcc lattice of the pure Ca can be described as a rombohedral
lattice with lattice parameter 3.95 ${\rm \AA}$ and angle $60^o$.
Note that the C$_6$Ca crystal structure is not equivalent to that
reported in \cite{Weller} which has a stacking $\alpha\beta$. In \cite{Weller} the 
structure determination was probably affected by the non-bulk character of the
 samples. 

Density Functional Theory (DFT) calculations are performed using the 
PWSCF/espresso code\cite{PWSCF} and
the generalized gradient approximation
(GGA) \cite{PBE}. We use ultrasoft pseudopotentials\cite{Vanderbilt}
with valence configurations 3s$^2$3p$^6$4s$^2$ for Ca and
2s$^2$2p$^2$ for C. The electronic
wavefunctions and the charge density are expanded using a 30 
and a 300 Ryd cutoff. 
The dynamical matrices and the electron-phonon coupling are calculated using
Density Functional Perturbation Theory in the linear response\cite{PWSCF}.
For the electronic integration in the phonon calculation we use
a $N_{k}=6\times6\times6$ uniform k-point mesh\cite{footnotemesh} and
and Hermite-Gaussian smearing of 0.1 Ryd.
For the calculation of
the electron-phonon coupling and of the electronic density of states
(DOS) we use a finer $N_k=20\times 20\times 20$ mesh.
For the $\lambda$ average over the phonon momentum {\bf q} 
we use a  $N_q=4^3$ ${\bf q}-$points mesh.
The phonon dispersion is obtained by Fourier interpolation of the 
dynamical matrices computed on the $N_q$ points mesh. 

The DFT band structure is shown in figure \ref{fig:bands}(b).
Note that the $\Gamma\chi$X direction and the L$\Gamma$ direction are
parallel and perpendicular to the graphene layers. 
The K special point of the graphite lattice is refolded at $\Gamma$ 
in this structure.
For comparison we plot in \ref{fig:bands}(c) 
the band structure of C$_{6}$Ca and with Ca atoms removed 
(C$_6$$^{*}$) and the structure C$_{6}$Ca with C$_6$ atoms 
removed ($^{*}$Ca). The size of the red dots in fig. \ref{fig:bands}(b)
represents the percentage of Ca component in a given band (L\"owdin population). 
The $^{*}$Ca band has a free electron like dispersion as
in fcc Ca.
From the magnitude of the Ca component and from the comparison between 
fig. \ref{fig:bands}(b) and (c) we conclude
that the C$_6$Ca bands can be interpreted as a superposition of the
$^{*}$Ca and of the C$_6$$^{*}$ bands.
At the Fermi level, one band 
originates from the free electron like $^{*}$Ca band and
disperses in all the directions.
The other bands correspond to the $\pi$ bands in C$_6$$^{*}$
and are weakly dispersive in the direction perpendicular to the 
graphene layers.
The Ca band has been incorrectly interpreted as 
an interlayer-band \cite{Csanyi} not associated to metal orbitals. 

More insight on the electronic states at E$_f$ can be obtained calculating
the electronic DOS.
The total DOS, fig. \ref{fig:bands}(a),
is in agreement
with the one of ref. \cite{Mazin} and
at E$_f$ it is 
$N(0)=1.50$ states/(eV unit cell). 
We also report in fig. \ref{fig:bands}(a) the atomic-projected density of state using the L\"owdin populations,
$\rho_{\eta}(\epsilon)=\frac{1}{N_k}\sum_{{\bf k}n}|\langle \phi^{L}_{\eta}|\psi_{{\bf k}n}\rangle|^2
\delta(\epsilon_{{\bf k}n}-\epsilon)$. In this expression 
$|\phi^{L}_{\eta}\rangle=\sum_{\eta\prime}[{\bf S}^{-1/2}]_{\eta,\eta^{\prime}} |\phi^{a}_{\eta^{\prime}
}\rangle$ are the orthonormalized L\"owdin orbitals, $ |\phi^{a}_{\eta^{\prime}}\rangle$ are the
atomic wavefunctions and
$ S_{\eta,\eta^{\prime}}=\langle \phi^{a}_{\eta} |\phi^{a}_{\eta^{\prime}}\rangle$.
The Kohn and Sham energy bands and wavefunctions 
are $\epsilon_{{\bf k}n}$ and $|\psi_{{\bf k}n}\rangle$.
This definition leads to projected DOS which are
unambiguously determined and are independent
of the method used for the electronic structure calculation.
At E$_f$ the Ca 4s, Ca 3d, Ca 4p, C 2s, C 2p$_{\sigma}$ and C 2p$_{\pi}$ are 
0.124, 0.368, 0.086, 0.019, 0.003, 0.860 states/(cell eV), respectively.
Most of C DOS at E$_f$ comes from C  2p$_{\pi}$ orbitals.
Since the sum of all the projected DOSs is almost identical to the 
total DOS, the electronic states at E$_f$ are very well described by
a superposition of atomic orbitals. Thus 
the occurrence of a non-atomic interlayer-state, proposed in 
ref. \cite{Csanyi}, is further excluded.
From the integral of the projected DOSs we
obtain a charge transfer of 0.32 electrons (per unit cell)
to the Graphite layers ($\Delta=0.11$). 
\begin{figure}[t]
\includegraphics[width=0.9\columnwidth]{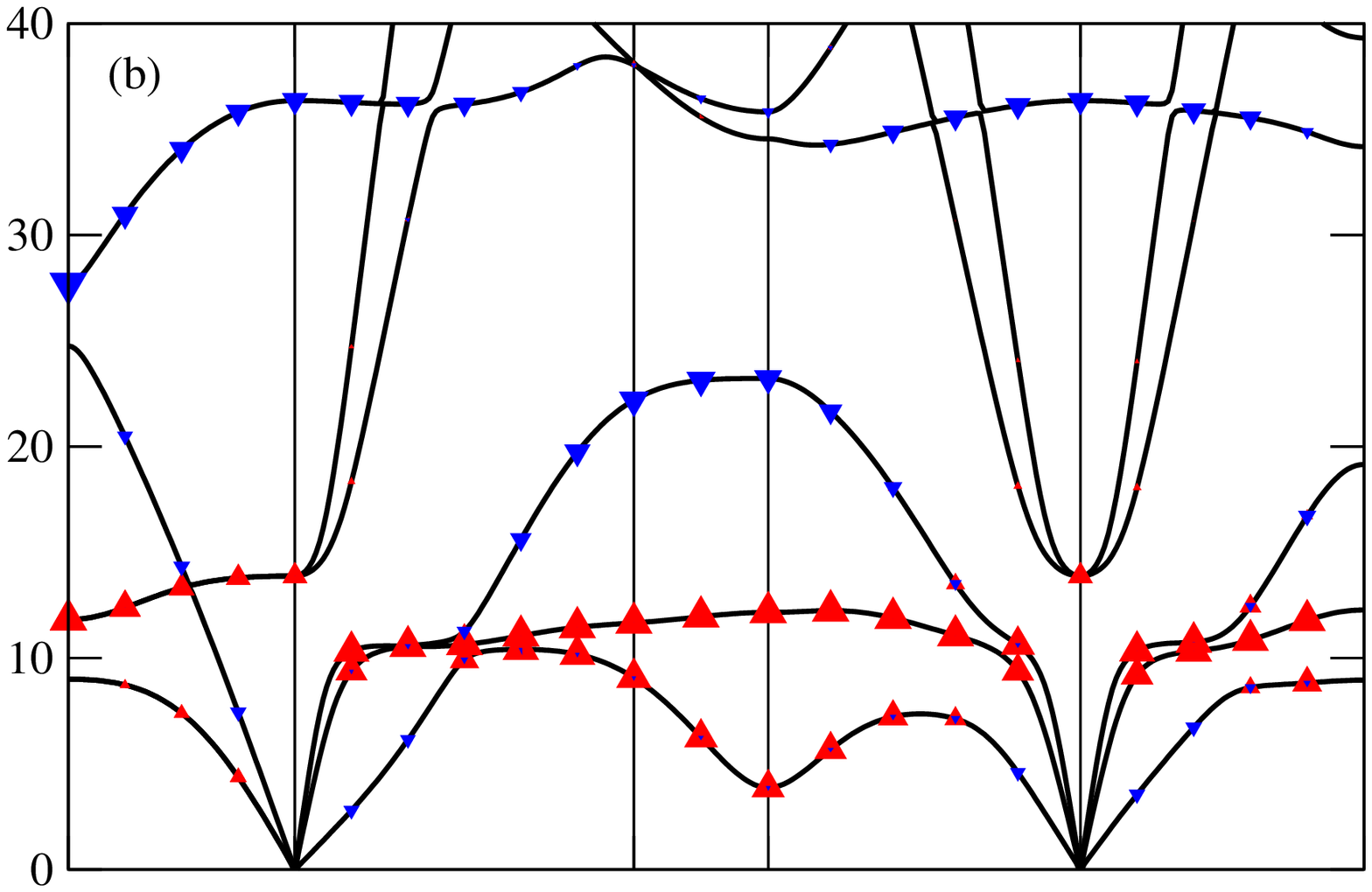}
\includegraphics[width=0.9\columnwidth]{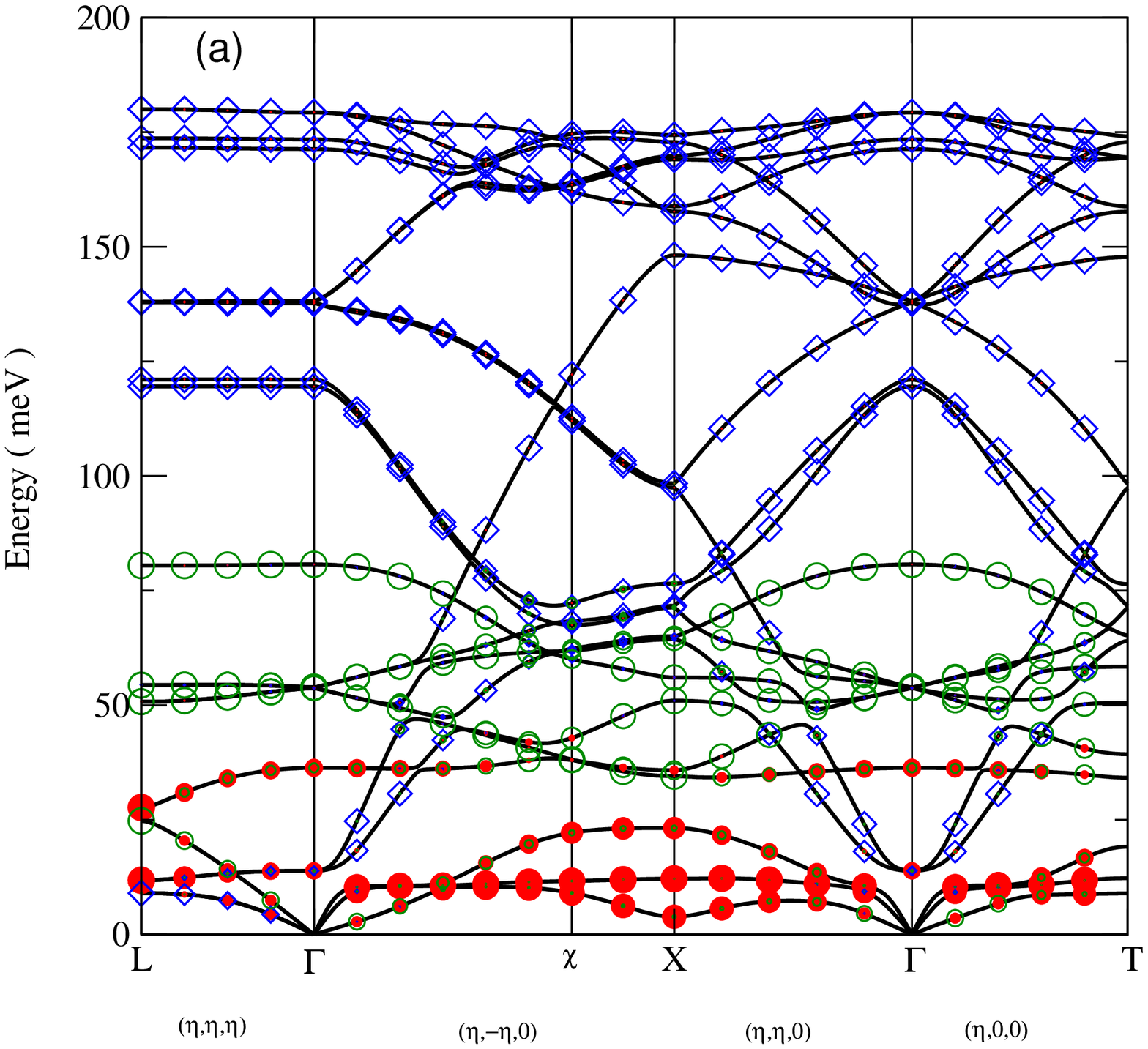}
\caption{(Color online) (a) and (b) CaC$_6$ Phonon dispersion. 
The amount of Ca vibration is indicated by the size of the 
\textbullet, of C$_z$  by the size of $\circ$, 
of C${xy}$  by the size of $\diamond$, of
Ca$_{xy}$ by the size of $\blacktriangle$ and of
Ca$_z$ by the size of $\blacktriangledown$.}
\label{fig:branchie}
\end{figure}

\begin{figure}[t]
\includegraphics[width=\columnwidth]{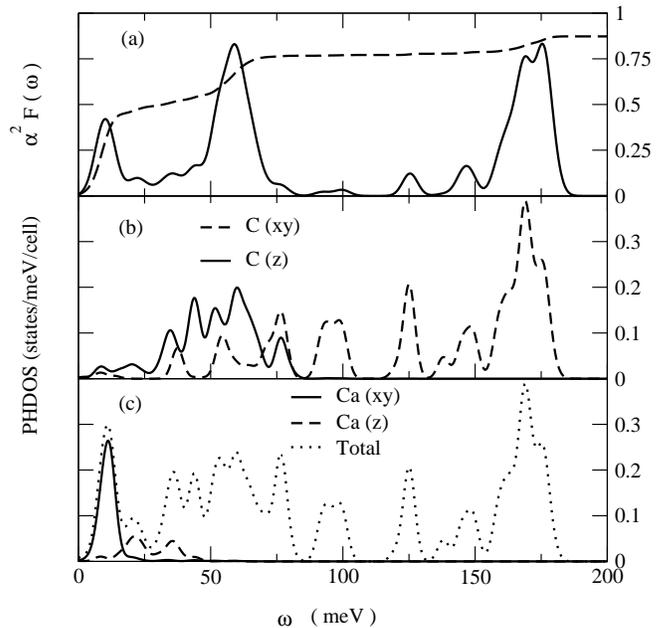}
\caption{(a) Eliashberg function, $\alpha^2F(\omega)$, (continuous line) and
integrated coupling, $\lambda(\omega)$ (dashed). (b) and (c) PHDOS projected on 
selected vibrations and total PHDOS.}
\label{fig:alpha2f}
\end{figure}

The phonon dispersion ($\omega_{{\bf q}\nu}$) is shown in fig. \ref{fig:branchie}. 
For a given mode $\nu$ and at a given momentum ${\bf q}$, the radii
of the symbols in fig.\ref{fig:branchie} indicate the 
square modulus of the displacement decomposed in Ca and C in-plane ($xy$, parallel to the
graphene layer) and out-of-plane ($z$, perpendicular to the graphene layer) contributions. 
The corresponding phonon density of states (PHDOS) are shown in fig. \ref{fig:alpha2f}
(b) and (c).
The decomposed PHDOS are well separated in energy. 
The graphite modes are weakly dispersing in the out-of-plane direction while
the Ca modes are three dimensional. However the Ca$_{xy}$ and the Ca$_z$ vibration
are well separated contrary to what expected for a perfect fcc-lattice.
One Ca$_{xy}$ vibration is an Einstein mode being weakly dispersive in all directions.

The superconducting properties of C$_6$Ca can be understood calculating
the electron-phonon interaction for a phonon mode $\nu$ with momentum ${\bf q}$:
\begin{equation}\label{eq:elph}
\lambda_{{\bf q}\nu} = \frac{4}{\omega_{{\bf q}\nu}N(0) N_{k}} \sum_{{\bf k},n,m} 
|g_{{\bf k}n,{\bf k+q}m}^{\nu}|^2 \delta(\epsilon_{{\bf k}n}) \delta(\epsilon_{{\bf k+q}m})
\end{equation}
where the sum is over the Brillouin Zone.
The matrix element is
$g_{{\bf k}n,{\bf k+q}m}^{\nu}= \langle {\bf k}n|\delta V/\delta u_{{\bf q}\nu} |{\bf k+q} m\rangle /\sqrt{2 \omega_{{\bf q}\nu}}$,
where $u_{{\bf q}\nu}$ is the amplitude of the displacement of the phonon 
and $V$ is the Kohn-Sham potential.
The electron-phonon coupling is  
$\lambda=\sum_{{\bf q}\nu} \lambda_{{\bf q}\nu}/N_q = 0.83$.
We show in fig.\ref{fig:alpha2f} (a) the Eliashberg function
\begin{equation}
\alpha^2F(\omega)=\frac{1}{2 N_q}\sum_{{\bf q}\nu} \lambda_{{\bf q}\nu} \omega_{{\bf q}\nu} \delta(\omega-\omega_{{\bf q}\nu} )
\end{equation}
and the integral $\lambda(\omega)=2 \int_{-\infty}^{\omega} d\omega^{\prime} 
\alpha^2F(\omega^{\prime})/\omega^{\prime}$. Three main contributions to $\lambda$ can 
be identified associated to Ca$_{xy}$, C$_z$ and C$_{xy}$ vibrations. 
A more precise estimate of the different contributions can be obtained
noting that
\begin{equation}\label{eq:trlambda}
\lambda=
\frac{1}{N_q}\sum_{\bf q}
\sum_{i\alpha j\beta} [{\bf G}_{\bf q}]_{i\alpha,j\beta} [{\bf C_q}^{-1}]_{j\beta,i\alpha}
\end{equation}
where $i,\alpha$ indexes indicate the displacement in the Cartesian direction $\alpha$
of the $i^{\rm th}$ atom,  
$[{\bf G_q}]_{i\alpha,j\beta}=\sum_{{\bf k},n,m}4 {\tilde g}_{i\alpha}^{*}{\tilde g}_{j\beta}
\delta(\epsilon_{{\bf k}n}) \delta(\epsilon_{{\bf k+q}m})/[N(0) N_{k}]$, and 
${\tilde g}_{i\alpha}=\langle {\bf k}n|\delta V/\delta x_{{\bf q} i\alpha} |{\bf k+q} m\rangle
 /\sqrt{2}$. 
The ${\bf C_q}$ matrix is the Fourier transform of the force constant matrix 
(the derivative of the forces respect to the atomic displacements).
We decompose $\lambda$ restricting the summation over $i,\alpha$ and that over
$i,\beta$ on two sets of atoms and Cartesian directions. The sets are
C$_{xy}$, C$_{z}$, Ca$_{xy}$, and Ca$_z$. The resulting $\bm{\lambda}$ matrix is:
\begin{equation}
  \bm{\lambda}\,=
  \begin{matrix}
    &
    \begin{matrix}
       {\rm C}_{xy} &  {\rm C}_{z} & {\rm Ca}_{xy} & {\rm Ca}_z \\
    \end{matrix} \\
    \begin{matrix}
       {\rm C}_{xy} \\
       {\rm C}_{z}  \\
       {\rm Ca}_{xy}\\
       {\rm Ca}_z   \\
    \end{matrix} &
    \begin{pmatrix}
       0.12 & 0.00 & 0.00 & 0.00 \\
       0.00 & 0.33 & 0.04 & 0.01 \\
       0.00 & 0.04 & 0.27 & 0.00 \\
       0.00 & 0.01 & 0.00 & 0.06 \\
    \end{pmatrix}
  \end{matrix}
\end{equation}

The off-diagonal elements are negligible. The Ca out-of-plane and C in-plane
contributions are small.
For the in-plane C displacements, eq. \ref{eq:model} with
$\Delta=0.11$ gives $\lambda_{{\rm C}_{xy},{\rm C}_{xy}}=0.11$. Such
a good agreement is probably fortuitous given the oversimplified assumptions
of the model. The main contributions
to $\lambda$ come from Ca in-plane and C out-of-plane displacements.
As we noted previously the C out-of-plane vibration do not couple with 
the C $\pi$ Fermi surfaces. Thus the coupling to the C out-of-plane
displacements comes from electrons belonging to the Ca Fermi surface. 
Contrary to what expected in an fcc lattice, 
the Ca$_{xy}$ phonon frequencies are smaller than the  Ca$_{z}$ ones. This can be
explained from the much larger $\lambda$ of the Ca in-plane modes.

The critical superconducting temperature is estimated using the McMillan 
formula\cite{mcmillan}:
\begin{equation}
T_c = \frac{\langle \omega \rangle}{1.2} \exp\left( - \frac{1.04 (1+\lambda)}{\lambda-\mu^* (1+0.62\lambda)}\right)\label{eq:mcmillan}
\end{equation}
where $\mu^*$ is the screened Coulomb pseudopotential
and $\langle\omega\rangle=24.7$ meV is the phonon frequencies 
logarithmic average. We obtain T$_c=11$K, with  $\mu^{*}=0.14$.
We calculate the isotope effect 
by neglecting the dependence of $\mu^{*}$ on $\omega$.
We calculate the parameter $\alpha({\rm X})=-\frac{d \log{T_c}}{d M_{\rm X}}$ 
where X is C or Ca. We get $\alpha({\rm Ca})=0.24$ and $\alpha({\rm C})=0.26$.
Our computed $\alpha({\rm Ca})$ is substantially smaller
than the estimate given in ref. \cite{Mazin}. This is due to the fact that 
only $\approx 40\%$ of $\lambda$ comes from the coupling to Ca phonon modes and not 
$85\%$ as stated in ref.\cite{Mazin}. 

In this work we have shown that superconductivity in C$_6$Ca is due to
an electron-phonon mechanism. The carriers are mostly electrons in the Ca Fermi 
surface coupled with Ca in-plane and C out-of-plane phonons.  
Coupling to both modes is important, as can be easily inferred from the
calculated isotope exponents $\alpha({\rm Ca})=0.24$ and $\alpha({\rm C})=0.26$.
Our results suggest a general mechanism for the
occurrence of superconductivity in GICs. In order to stabilize a 
superconducting state it is necessary to have an intercalant Fermi surface 
since the simple doping of the $\pi$ bands in graphite does not lead to a 
sizeable electron-phonon coupling.
This condition occurs if the intercalant band is partially occupied,
i. e. when the intercalant is not fully ionized.
The role played in superconducting GICs by the intercalant 
Fermi surface has been previously suggested by \cite{Jishi}.
More recently a correlation between the presence of a band, not belonging
to graphite, and superconductivity has been observed in \cite{Csanyi}. 
However the attribution of this band to an interlayer state not derived
from intercalant atomic orbitals is incorrect.

We acknowledge illuminating discussions with M. Lazzeri,G. Loupias,
M. d'Astuto, C. Herold and A. Gauzzi.  Calculations
were performed at the IDRIS supercomputing center (project 051202).

\end{document}